\documentclass[twocolumn]{aa}
\usepackage{graphicx}
\usepackage{txfonts}
\begin{document}
\title{Strong effects of time-dependent ionization in early SN~1987A }

\author{Victor P. Utrobin\inst{1,2} \and Nikolai N. Chugai\inst{3}}

\offprints{V. Utrobin, \email{utrobin@itep.ru}}

\institute{
   Max-Planck-Institut f\"ur Astrophysik,
   Karl-Schwarzschild-Str. 1, D-85741 Garching, Germany
\and
   Institute of Theoretical and Experimental Physics,
   B.~Cheremushkinskaya St. 25, 117259 Moscow, Russia
\and
   Institute of Astronomy of Russian Academy of Sciences,
   Pyatnitskaya St. 48, 109017 Moscow, Russia}

\date{Received 23 December 2004 / accepted 16 March 2005}

\abstract{
We study a time-dependent hydrogen ionization in the atmosphere of SN~1987A
   during the first month after the explosion.
The model includes kinetics of hydrogen ionization and excitation, molecular
   hydrogen kinetics, and a time-dependent energy balance.
The primary strong effect of the time-dependent ionization is the enhanced
   hydrogen ionization compared to the steady-state model.
The time-dependent ionization provides a sufficient population of excited
   hydrogen levels to account for the observed H$\alpha$ without invoking
   the external $^{56}$Ni.
We find that the Ba II 6142 \AA\ line in SN~1987A can be reproduced for
   the LMC barium abundance.
This resolves the long-standing problem of the unacceptably high barium
   overabundance in SN~1987A.
The key missing factor that should be blamed for the "barium problem" is
   the time-dependent ionization.
The modelling of the H$\alpha$ profile on day 4.64 indicates 
   the ratio of the kinetic energy to the ejected mass
   $\approx 0.83\times 10^{50} \mathrm{erg}\,M_{\sun}^{-1}$.

}
%
\titlerunning{Time-dependent effects in SN~1987A}
\authorrunning{V. P. Utrobin \& N. N. Chugai}
\maketitle

\section{Introduction}

Spectra of type IIP supernovae (SN) combined with photometric data are
   the primary source of our knowledge about the mass, energy, chemical
   composition, mixing, asymmetry of ejecta that are of a vital importance
   for the verification of the explosion mechanism.
Generally, to recover the information imprinted in the spectra one needs
   to use rather elaborated models.
Unfortunately, the ultimate model with all the physics included is beyond
   reach, so some simplifications are unavoidable.
One of the assumptions accepted in the atmosphere models is the steady-state
   approximation for the ionization kinetics.

Yet, that the time-dependent effect of the hydrogen ionization may play
   a role in the SN envelope has been recognized by Kirshner \& Kwan
   (\cite{kk75}); they applied this effect to account for the high H$\alpha$
   luminosity of the type II SN~1970G.
The time-dependent ionization has been exploited for SN~1987A to account for
   the high excitation of hydrogen in the outer atmosphere
   ($v>7000$ km\,s$^{-1}$) during the first 40 days (Chugai \cite{chugai91a}).
The crucial role of the time-dependent ionization at the late nebular epoch
   of SN~1987A has been emphasized by Clayton et al. (\cite{cltjk92}) and
   Fransson \& Kozma (\cite{fk93}).

Recently, using a time-dependent model of the hydrogen ionization with
   the kinetics of atomic and molecular hydrogen we have found that
   time-dependent effects in SN~1987A were strong during the first month
   and that they could provide the hydrogen excitation required to account for
   the H$\alpha$ line strength (Utrobin \& Chugai \cite{uc02}) which was
   underproduced in steady-state models of SN~1987A atmosphere.
However, our previous model has ignored a time-dependent energy balance;
   instead simple laws have been used to mimic the evolution of the electron
   temperature in the atmosphere.

Keeping in mind the importance of the non-steady state ionization effects
   for the understanding of the phenomena seen in early SN~1987A, we revisit
   this issue.
To this end we upgrade the atmosphere model that apart from the time-dependent
   kinetics includes also the time-dependent energy balance.
Our main goal is to compute the ionization and excitation of hydrogen in
  the atmosphere of SN~1987A for different epochs, and to compare the
  calculated H$\alpha$ line with the observations.
We also address a highly intriguing problem of tremendous Ba overabundance
   that has been recovered by several studies (Mazzali et al. \cite{mlb92}; 
   and references there).
Our preliminary guess was that the Ba overabundance problem was caused by
   the strong underestimation of the Ba II ionization fraction.

We begin with a question, why time-dependent effects are important
   (Sect.~\ref{sec:general}) and then describe in detail the proposed
   model (Sect.~\ref{sec:model}).
The H$\alpha$ strength on day 4.64 is used to constrain the kinetic energy
   of the hydrodynamic model Sect.~\ref{sec:res-hahyd}, the computations of
   the ionization and temperature structure of the atmosphere along with
   the H$\alpha$ profiles for selected phases are presented in
   Sect.~\ref{sec:res-h}, while in Sect.~\ref{sec:res-ba2} we address the
   problem of high Ba overabundance.

We adopt here the radial velocity of SN 1987A of 286.5 km\,s$^{-1}$
   (Meaburn et al. \cite{mbh95}) and the distance of 50 kpc.

\section{Why time-dependent?}
\label{sec:general}

A widely shared view that the steady-state statistical equilibrium is
   a reasonable approximation for SN~IIP atmospheres relies on the claim
   that the recombination time is much smaller than the SN age
\begin{equation}
   (\alpha N_{\mathrm{e}})^{-1} \ll t \; ,
   \label{eq:trec}
\end{equation}
   where $\alpha$ is the coefficient of recombination to the excited states,
   $N_{\mathrm{e}}$ is the electron number density.
Adopting $\alpha \sim 3 \times 10^{-13}$ cm$^3$\,s$^{-1}$ and a typical SN age
   at the photospheric epoch $t \sim 10^6-10^7$ s one finds that the
   requirement (\ref{eq:trec}) is satisfied for $N_{\mathrm{e}} \gg
   10^7$ cm$^{-3}$.
This condition is actually met at the photosphere thus supporting the
   steady-state approximation.
Meanwhile, it is violated in the outer, high velocity layers,
   where $N_{\mathrm{e}}$ may drop as low as $10^7$ cm$^{-3}$.
Therefore, generally, the time dependent approach is needed at least,
   if one addresses the situation in the outer layers.

Moreover, it turns out that the steady-state approximation is violated
   also in the inner atmosphere, where the condition (\ref{eq:trec}) holds.
The point is that this inequality has little in common with the condition of
   the steady state for multilevel atom irradiated by the optical continuum,
   because of the efficient reionization of recombined atom.
The importance of this phenomenon has been first recognized for the problem
   of the hydrogen recombination in the early universe, where L$\alpha$
   trapping along with the reionization from the second level results in
   the significant increase of the recombination time
   $t_{\mathrm{rec}}=-N_{\mathrm{e}}/(dN_{\mathrm{e}}/dt)$ 
   (Zeldovich et al. \cite{zks68}).

The net recombination rate $dN_{\mathrm{e}}/dt$ refers to transitions 
   to the ground state that either do not emit L$\alpha$ or followed 
   by the escape of the emitted L$\alpha$ quantum without resonance scattering.
Let the probability of such a transition is $w_{21}$.
Then in the "two level plus continuum" approximation the net recombination
   rate (neglecting the expansion effect) is
\begin{equation}
   \frac{dN_{\mathrm{e}}}{dt} = - \alpha N_{\mathrm{e}}^2 w_{21} \; .
   \label{eq:chtrec}
\end{equation}
With all the major processes from the second level taken into account
   the probability $w_{21}$ is
\begin{equation}
   w_{21} = \frac{A_{\mathrm{2q}} + N_{\mathrm{e}} q_{21} + A_{21} \beta_{21}}
            {P_2 + A_{\mathrm{2q}} + N_{\mathrm{e}} q_{21} + A_{21} \beta_{21}} \; ,
   \label{eq:w21}
\end{equation}
   where
   $P_2$ is the photoionization rate from the second level;
   $A_{\mathrm{2q}}$ is the probability of the two-photon decay evaluated
   for the equipartition of 2s and 2p levels;
   $q_{21}$ is the collisional deexcitation coefficient;
   $A_{21} \beta_{21}$ is the L$\alpha$ escape probability.

Let us estimate $w_{21}$ for the neutral hydrogen number density
   $N_1>10^9$ cm$^{-3}$ and the electron number density $N_{\mathrm{e}}<10^8$
   cm$^{-3}$, when the two-photon transition dominates over collisional
   transition and L$\alpha$ escape, i.e., $w_{21} \approx A_{\mathrm{2q}}/P_2$.
This situation is typical for the inner atmosphere of SN~IIP.
With the photospheric temperature of $T \approx 5000$ K one gets
   $P_2 \approx 10^4 W$ s$^{-1}$, where $W$ is the dilution factor.
Close to the photosphere (for $W > 0.1$) one finds $w_{21} < 2 \times 10^{-3}$.
The effective recombination time is then
\begin{equation}
   t_{\mathrm{rec}} = - N_{\mathrm{e}}/(dN_{\mathrm{e}}/dt)
      = 1/(\alpha N_{\mathrm{e}} w_{21})
      \sim 500 /(\alpha N_{\mathrm{e}}) \; .
   \label{eq:treceff}
\end{equation}
This shows that the effective recombination time is by a factor of
   $1/w_{21} \sim 500$ greater than the recombination time to excited states
   $1/(\alpha N_{\mathrm{e}})$.
For $N_{\mathrm{e}} \sim 10^8$ cm$^{-3}$ one obtains
   $t_{\mathrm{rec}} > 10^7$~s $>t$.
We thus conclude that the steady-state approximation may break down in
   the dense inner atmosphere as well.

These qualitative considerations demonstrate that the hydrogen recombination
   in the atmosphere of SN~IIP at the photospheric epoch is essentially
   a time-dependent phenomenon.

\section{The model and input physics}
\label{sec:model}

A self-consistent treatment of the SN atmosphere requires the hydrodynamic
   modelling with time-dependent radiation transport, energy balance,
   and ionization kinetics.
At present such a general approach is beyond reach of computation possibilities.
To study time-dependent effects in the atmosphere of SN~1987A and yet to make
   the problem solvable, we sacrifice the radiation transfer in the atmosphere
   using, instead, a simple description of the continuum radiation field.

\subsection{Continuum radiation}
\label{sec:model-cont}
%
\begin{figure}[t]
   \resizebox{\hsize}{!}{\includegraphics{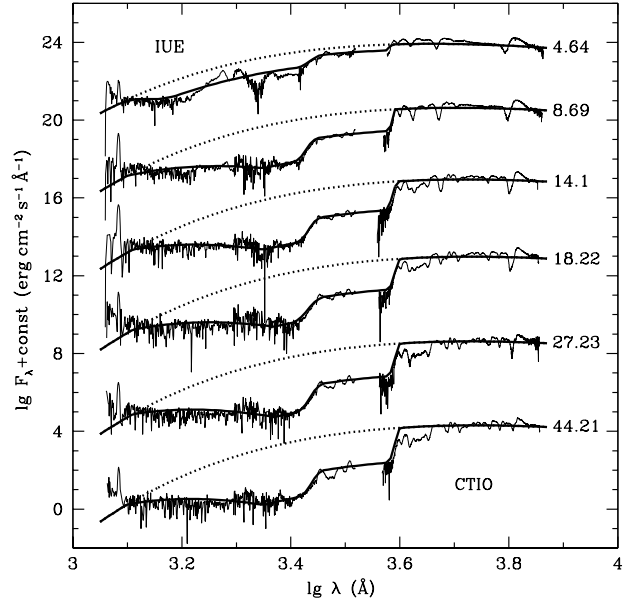}}
   \caption{%
   The combined UV (IUE) (Pun et al. \cite{pks95}) and optical (CTIO)
      (Phillips et al. \cite{phhn88}) spectra of SN~1987A (\emph{thin solid
      line\/}), the black-body continuum at the calculated effective temperature
      of the high-energy model (\emph{dotted line\/}), and the observed
      continuum (\emph{thick solid line\/}) between days 4.64 and 44.21.
   }
   \label{fig:flux}
\end{figure}
The SN atmosphere is very opaque in the Lyman continuum ($\nu > \nu_1$),
   so the photoionization by the photospheric Lyman continuum can be neglected.
On the other hand, the hydrogen ionization by the recombination Lyman continuum
   in the atmosphere can be easily taken into account via cancelling the
   photoionization of hydrogen from the ground level and the recombination
   to the ground level in the rate equations.
In the atmosphere the radiation field in continuum at these frequencies is
   considered as being in the thermal equilibrium with matter.

The important simplification we adopt is that the continuum in the range of
   $\nu < \nu_1$ corresponds to the free-streaming case, i.e., the average
   intensity of the continuum in the atmosphere is $J_{\nu} = W I_{\nu}$,
   where $I_{\nu}$ is the specific intensity of the photospheric radiation.
The photospheric radius $R_\mathrm{ph}$ and effective temperature
   $T_\mathrm{eff}$ are provided by the hydrodynamic model.

The free-streaming approximation is a good approximation in the visual band,
   where the optical depth of the atmosphere is quite low.
However, for the ultraviolet (UV) radiation between Lyman and Balmer edges
   this approximation is indeed rather artificial, because of the large optical
   depth due to the UV line opacity.
As a result, the average intensity of the UV radiation in the atmosphere
   generally should differ from the free-streaming approximation.

Nevertheless, this approximation is not as bad as one would think at first
   glance.
It predicts behavior of $J \propto r^{-2}$ in the most of the atmosphere.
In the case of pure scattering with the constant scattering coefficient
   ($\chi(r)=const$) one obtains for the extended atmosphere
   $J \propto [1+1.5R\chi(R/r-1)]$ where $R$ is the outer radius of
   the atmosphere (Chandrasekhar \cite{chandra34}).
According to this expression the behavior of $J(r)$ is steeper than $r^{-2}$
   for outer layers ($r \sim R$) and $J(r) \propto r^{-1}$ for $r \ll R$.
So, our approximation corresponds to some intermediate range in the atmosphere.
Moreover, since the scattering of UV metal lines is liable to photon splitting,
   the interaction of the UV radiation with matter will be close to the
   absorption rather than scattering.
We expect, therefore, more steep behavior of $J(r)$ in the inner layer
   compared to the pure scattering case.
In that case the law $J \propto r^{-2}$ may be a sensible approximation
   in the deeper layers as well.

To further allow for the uncertainty of our approximation we consider
   two extreme cases: (i) the photospheric brightness is black-body with
   the effective temperature (model A); (ii) the photospheric brightness
   corresponds to the observed spectrum (model B)  (Fig.~\ref{fig:flux}).
The real situation is somewhere between these two cases.

\subsection{Hydrodynamic model}
\label{sec:model-hydro}
%
\begin{figure}[t]
   \resizebox{\hsize}{!}{\includegraphics{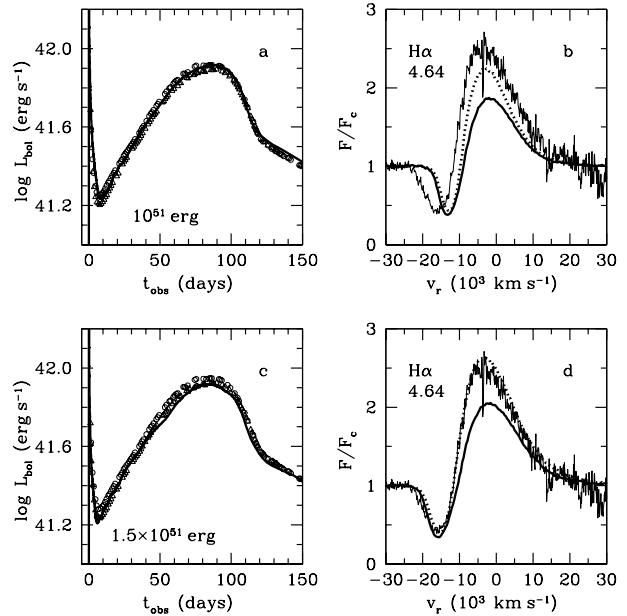}}
   \caption{%
   Comparison of the low and high-energy models of SN~1987A.
   Top: (\textbf{a}) bolometric light curve of the low-energy model
      (\emph{thick solid line\/}) compared with the observations of SN~1987A
      obtained by Catchpole et al. (\cite{cmm87}, \cite{cwf88})
      (\emph{open circles\/}) and Hamuy et al. (\cite{hsgm88})
      (\emph{open triangles\/}), and (\textbf{b}) the H$\alpha$ line profile
      on day 4.64 computed on the basis of this hydrodynamic model for the
      black-body continuum (model A) (\emph{dotted line\/}) and the observed
      continuum (model B) (\emph{thick solid line\/}). The calculated spectra
      are overplotted on the observed spectrum (Phillips et al. \cite{phhn88})
      (\emph{thin solid line\/}).
   Bottom: (\textbf{c}) bolometric light curve of the high-energy model and
      (\textbf{d}) the corresponding H$\alpha$ profile on day 4.64.
   }
   \label{fig:hydmods}
\end{figure}
The density distribution, chemical composition, radius of the
   photosphere and effective temperature are provided by the hydrodynamic
   model of SN~1987A (see for details Utrobin \cite{utrobin04}).
We consider the non-evolutionary model N of Utrobin (\cite{utrobin04}) with
   18 $M_{\sun}$ ejecta, kinetic energy of $10^{51}$ erg, and $^{56}$Ni mass
   of 0.073 $M_{\sun}$.
For the distance modulus to the LMC of $m-M=18.5$ mag, color excess of
   $E(B-V)=0.15$ mag, and an interstellar extinction of $A_V=3.1E(B-V)$ mag
   a progenitor of SN~1987A, Sanduleak $-69^{\circ}202$, had the radius
   of 46.8 $R_{\sun}$.
To produce the presupernova density distribution the evolutionary model l20n2ae
   of Woosley et al. (\cite{whwl97}) was rescaled to adopted radius and ejecta
   mass.
We assumed for the outer layers the circumstellar helium abundance $Y=0.441$
   (Wang \cite{wang91}; Lundqvist \& Fransson \cite{lf96}) and
   the LMC metallicity $Z=0.004$ (Dufour \cite{dufour84}).
The structure of the presupernova outer layers was modified to increase
   density compared with the evolutionary model of non-rotating single star.
This correction was needed to fit the observed bolometric light curve of
   SN~1987A (Fig.~\ref{fig:hydmods}a) with a moderate mixing of radioactive
   $^{56}$Ni in the velocity range $\le 2500$ km\,s$^{-1}$ (Utrobin
   \cite{utrobin04}).
The final model is dubbed as the low-energy model.

We consider also the high-energy version of the above model.
This model with the kinetic energy $1.5\times10^{51}$ erg differs from the
   low-energy model by the presupernova radius of 35.0 $R_{\sun}$, $^{56}$Ni
   mass of 0.075 $M_{\sun}$, and the adopted color excess of $E(B-V)=0.17$ mag
   (Michael et al. \cite{mmc03}).
The chemical composition of the outer layers is typical to the LMC chemical
   composition, $X=0.743$, $Y=0.251$, and $Z=0.006$ (Dufour \cite{dufour84}).
This model agrees with the observed bolometric light curve of SN~1987A for
   $^{56}$Ni mixed in the velocity range $\le 3000$ km\,s$^{-1}$
   (Fig.~\ref{fig:hydmods}c).

\subsection{Line transfer}
\label{sec:model-line}

Starting the first day the envelope of SN~1987A expands homologously
   (Utrobin \cite{utrobin04}), so the Sobolev approximation is applicable.
In this approximation the line transfer is described by the escape probability
   (Sobolev \cite{sobolev60}; Castor \cite{castor70})
\begin{equation}
   \beta_{ul} = \frac{1 - exp(-\tau_{lu})}{\tau_{lu}} \; ,
   \label{eq:beta}
\end{equation}
   where
\begin{equation}
   \tau_{lu} = \frac{\pi e^2}{m_\mathrm{e} c} f_{lu} \lambda_{lu} t
            \left( N_l - \frac{g_l}{g_u} N_u \right)
   \label{eq:taulu}
\end{equation}
   is the Sobolev optical depth in a transition from level $l$ to level $u$.
Here $N_l$, $g_l$, $N_u$, $g_u$ are number densities and statistical weights
   of atoms in the lower and upper levels, respectively, $f_{lu}$ is the
   oscillator strength, and $\lambda_{lu}$ is the line wavelength.
In the Sobolev approximation the frequency-averaged mean intensity
   of the line is
\begin{equation}
   J_{lu} = (1 - \beta_{ul}) S_{lu} + \beta_{ul} J(\nu_{lu}) \; ,
   \label{eq:jlu}
\end{equation}
   where
\begin{equation}
   S_{lu} = \frac{2 h \nu_{lu}^3}{c^2}
         \left( \frac{g_u N_l}{g_l N_u} - 1 \right)^{-1}
   \label{eq:slu}
\end{equation}
   is the line source function and $\nu_{ul}$ is the line frequency.
The net frequency-integrated line emissivity is
\begin{equation}
   \eta_{ul}= \frac{1}{4\pi} h \nu_{lu} N_u A_{ul} \beta_{ul}
             \left( 1 - \frac{J(\nu_{lu})}{S_{lu}} \right) ,
   \label{eq:etaul}
\end{equation}
   where $A_{ul}$ is the Einstein spontaneous emission probability for
   the transition from level $u$ to level $l$.

For the very opaque L$\alpha$ line the Sobolev approximation is violated,
   if one consider the L$\alpha$ radiation transfer, but remains valid,
   if one is interested in the computations of the escape rate
   (Chugai \cite{chugai80}).

\subsection{Gamma-rays and positrons deposition}
\label{sec:model-gdep}

The gamma rays with energy of about 1 MeV from the decay chain
   $^{56}$Ni $\to ^{56}$Co $\to ^{56}$Fe deposit their energy through
   Compton scattering by free and bound electrons.
The Compton electrons lose their energy through the Coulomb heating of free
   electrons and the ionization and excitation of atoms and ions.
The rates of the heating, non-thermal ionization and excitation for atoms and
   ions are taken according to Kozma \& Fransson (\cite{kf92}).
The gamma-rays transport is calculated in the approximation of the effective
   absorption opacity of 0.06 $Y_{\mathrm e}$ cm$^2$\,g$^{-1}$, where
   $Y_{\mathrm e}$ is the number of electrons per baryon.
Positrons are assumed to deposit their energy locally.

\subsection{Rate equations}
\label{sec:model-rteq}
The following elements and molecules are calculated in time-dependent non-LTE
   chemical kinetics: H, He, C, N, O, Ne, Na, Mg, Si, S, Ar, Ca, Fe, Ba,
   H$^{-}$, H$_{2}$, H$_{2}^{+}$, and H$_{3}^{+}$.
All elements but H are treated with the three ionization stages.
Neutral hydrogen and ions Mg II, Fe II, and Ba II are modelled with 15, 9,
   30, and 17 levels, respectively, while other atoms and ions are assumed
   to consist of the ground state and continuum.
The reaction network includes all bound-bound and bound-free, radiative and
   collisional processes for the atoms and ions with a detailed level structure
   (Appendices~\ref{apx:atmdt-hydr}--\ref{apx:atmdt-barium}),
   all bound-free, radiative and collisional processes for the two-state atoms
   and ions (Appendix~\ref{apx:atmdt-other}).
We consider kinetics of hydrogen molecules that may affect the hydrogen
  ionization degree.
Seven radiative and 37 collisional processes for the molecules are taken
   into account (Appendices~\ref{apx:moldt-negat}--\ref{apx:moldt-other}).

For one-dimensional, spherically symmetric flow the equation of continuity
   simplifies to
\begin{equation}
   {\partial \rho \over \partial t} = - {\rho \over t_{exp}} \; , \,
   {1 \over t_{exp}} =
   \biggl( 2+\frac{\partial ln v}{\partial ln r} \biggr) {v \over r} \; ,
\label{eq:cont}
\end{equation}
   where ${\partial / \partial t}$ is the Lagrangean time derivative;
   $\rho$ is the density of the matter; $r$ and $v$ are radius and velocity,
   respectively.
The net rate of transitions between level $k$ and all other levels of neutral
   atom z by atomic processes is
\begin{eqnarray}
   \left[ {\partial N_{\mathrm{z}^0,k} \over \partial t} \right]
   & = &
   \sum\limits_{u>k} N_{\mathrm{z}^0,u} A_{uk}
   \left(1-\frac{J_{lu}}{S_{ku}}\right)
   + N_{\mathrm{z}^+} N_\mathrm{e} \alpha_k
   \nonumber \\
   & - &
   N_{\mathrm{z}^0,k} \sum\limits_{l<k} A_{kl}
   \left(1-\frac{J_{lu}}{S_{lk}}\right) - N_{\mathrm{z}^0,k} P_k
   \nonumber \\
   & + &
   N_\mathrm{e} \sum\limits_{u>k}
   \left( N_{\mathrm{z}^0,u} q_{uk} - N_{\mathrm{z}^0,k} q_{ku} \right)
   + N_{\mathrm{z}^+} N_\mathrm{e}^2 q_{c,k}
   \nonumber \\
   & - &
   N_\mathrm{e} \sum\limits_{l<k}
   \left( N_{\mathrm{z}^0,k} q_{kl} - N_{\mathrm{z}^0,l} q_{lk} \right)
   - N_\mathrm{e} N_{\mathrm{z}^0,k} q_{k,c}
   \nonumber \\
   & + &
   R_{k}^\gamma-R_{k,c}^\gamma \; ,
\label{eq:[dNz0kdt]}
\end{eqnarray}
   where
   $N_{\mathrm{z}^0,k}$ is the number density of neutral atom z at
   the level $k$;
   $N_{\mathrm{z}^+}$ is the number density of singly ionized element z;
   $N_\mathrm{e}$ is the number density of free electrons;
   $J_{lu}$ is the frequency-averaged mean intensity of the
   line~(\ref{eq:jlu});
   $S_{lu}$ is the line source function~(\ref{eq:slu});
   $\alpha_k$ and $P_k$ are the total radiative recombination
   coefficient, including the spontaneous and induced processes, and
   the photoionization rate of neutral atom z in level $k$,
   respectively (Mihalas \cite{mihalas78});
   $q_{lu}$ is the electron collisional excitation rate from level $l$ to
   level $u$;
   $q_{ul}=(N_{\mathrm{z}^0,l}/N_{\mathrm{z}^0,u})^*q_{lu}$ is the
   corresponding deexcitation rate and the ratio
   $(N_{\mathrm{z}^0,l}/N_{\mathrm{z}^0,u})^*$ for the local thermodynamic
   equilibrium (LTE) is given by the Boltzmann formula;
   $q_{k,c}$ is the electron collisional ionization rate from level $k$;
   $q_{c,k}=(N_{\mathrm{z}^0,k}/N_{\mathrm{z}^+}N_{\mathrm{e}})^*q_{k,c}$
   is the corresponding recombination coefficient and the LTE ratio
   $(N_{\mathrm{z}^0,k}/N_{\mathrm{z}^+}N_{\mathrm{e}})^*$ is given by the
   Saha equation;
   $R_{k}^\gamma$ and $R_{k,c}^\gamma$ are the non-thermal excitation and
   ionization rates of neutral atom z in level $k$, respectively
   (Kozma \& Fransson \cite{kf92}).

The rate equation for the number density $N_{\mathrm{H}^0,2}$ of neutral
   hydrogen in level $2$ is determined by the net rate of
   flow of particles into the unit volume with a characteristic time
   $t_{exp}$~(\ref{eq:cont}), the net rate~(\ref{eq:[dNz0kdt]}), hydrogen
   two-photon decays, and other depopulation processes:
\begin{eqnarray}
   {\partial N_{\mathrm{H}^0,2} \over \partial t}
   & = &
   - {N_{\mathrm{H}^0,2} \over t_{exp}}
   + \left[ {\partial N_{\mathrm{H}^0,2} \over \partial t} \right]
   - A_{2s,1s}^{2q} N_{\mathrm{H}^0,2s}
   \nonumber \\
   & - &
   k_{5} N_{\mathrm{H}^0} N_{\mathrm{H}^0,2s}
   - k_{6} N_{\mathrm{H}^0} N_{\mathrm{H}^0,2p}
   \nonumber \\
   & - &
   k_{8} N_{\mathrm{H}^0,1} N_{\mathrm{H}^0,2s}
   - k_{9} N_{\mathrm{H}^0,1} N_{\mathrm{H}^0,2p}
   \nonumber \\
   & - &
   k_{14} N_{\mathrm{H}_2} N_{\mathrm{H}^0,2}  \; ,
\label{eq:dN2dt}
\end{eqnarray}
   where
   $N_{\mathrm{H}^0,2s}=0.25 N_{\mathrm{H}^0,2}$ and $N_{\mathrm{H}^0,2p}=
   0.75 N_{\mathrm{H}^0,2}$ are the number density of hydrogen
   levels 2s and 2p calculated assuming only collisional transition
   between these states;
   $N_{\mathrm{H}^0}=\sum\limits_{k=1}^{15} N_{\mathrm{H}^0,k}$ is the number
   density of neutral hydrogen;
   $N_{\mathrm{H}_2}$ is the number density of H$_2$;
   $A_{2s,1s}^{2q}$ is the probability of the hydrogen two-photon decay;
   $k_5$, $k_6$, $k_8$, $k_9$, and $k_{14}$ are the rate coefficients for
   the corresponding reactions from Table~\ref{ListNNReac}.
The rate equations for the number density of neutral hydrogen in level $3$
   and higher levels $k>3$ are
\begin{eqnarray}
   {\partial N_{\mathrm{H}^0,3} \over \partial t}
   & = &
   - {N_{\mathrm{H}^0,3} \over t_{exp}}
   + \left[ {\partial N_{\mathrm{H}^0,3} \over \partial t} \right]
   - k_{7} N_{\mathrm{H}^0} N_{\mathrm{H}^0,3}
   \nonumber \\
   & - &
   k_{14} N_{\mathrm{H}_2} N_{\mathrm{H}^0,3}
\label{eq:dN3dt}
\end{eqnarray}
   and
\begin{equation}
   {\partial N_{\mathrm{H}^0,k} \over \partial t} =
   - {N_{\mathrm{H}^0,k} \over t_{exp}}
   + \left[ {\partial N_{\mathrm{H}^0,k} \over \partial t} \right]
   - k_{14} N_{\mathrm{H}_2} N_{\mathrm{H}^0,k}  \; ,
\label{eq:dNkdt}
\end{equation}
   respectively.
Similarly, the rate equation for the number density $N_{\mathrm{z}^{+},k}$ of
   ions $\mathrm{z}^{+}=\mathrm{Mg}^{+}, \mathrm{Fe}^{+}, \mathrm{Ba}^{+}$
   in level $k$ simply is
\begin{equation}
   {\partial N_{\mathrm{z}^{+},k} \over \partial t} =
   - {N_{\mathrm{z}^{+},k} \over t_{exp}}
   + \left[ {\partial N_{\mathrm{z}^{+},k} \over \partial t} \right] \; .
\label{eq:dz+kdt}
\end{equation}
The rate equations for ionized hydrogen, negative hydrogen ion, and molecular
   hydrogen H$_{2}$, H$_{2}^{+}$ and H$_{3}^{+}$ are given in
   Appendix~\ref{apx:rteqmolh}.

For the two-state atoms and ions the net rate of transitions, for example,
   from neutral atom z into ion $\mathrm{z}^{+}$ by ionization and
   recombination processes is given by
\begin{eqnarray}
   \left[ {\partial N_{\mathrm{z}^+} \over \partial t} \right]
   & = &
   N_{\mathrm{z}^0} (P_{\mathrm{z}^0} + N_\mathrm{e} q_{\mathrm{z}^0,c} +
   R_{\mathrm{z}^0,c}^\gamma)
   \nonumber \\
   & - &
   N_{\mathrm{z}^+} N_\mathrm{e} (\alpha_{\mathrm{z}^0} +
     N_\mathrm{e} q_{c,\mathrm{z}^0}) \; .
\label{eq:[dz+dt]}
\end{eqnarray}
Using the net rate~(\ref{eq:[dz+dt]}) the rate equations for elements with
   three ionization stages are easily generalized:
\begin{equation}
   {\partial N_{\mathrm{z}^+} \over \partial t} =
   - {N_{\mathrm{z}^+} \over t_{exp}}
   + \left[ {\partial N_{\mathrm{z}^+} \over \partial t} \right]
   - \left[ {\partial N_{\mathrm{z}^{2+}} \over \partial t} \right]
\label{eq:dz+dt}
\end{equation}
   and
\begin{equation}
   {\partial N_{\mathrm{z}^{2+}} \over \partial t} =
   - {N_{\mathrm{z}^{2+}} \over t_{exp}}
   + \left[ {\partial N_{\mathrm{z}^{2+}} \over \partial t} \right] \; .
\label{eq:dz2+dt}
\end{equation}
In particular, the rate equation for singly ionized oxygen taking account of
   the charge transfer processes in collisions of H$^{+}$ with O and
   O$^{+}$ with H is
\begin{eqnarray}
   {\partial N_{\mathrm{O}^+} \over \partial t}
   & = &
   - {N_{\mathrm{O}^+} \over t_{exp}}
   + \left[ {\partial N_{\mathrm{O}^+} \over \partial t} \right]
   - \left[ {\partial N_{\mathrm{O}^{2+}} \over \partial t} \right]
   \nonumber \\
   & + &
   k_{\mathrm{HO}} N_{\mathrm{H}^+} N_{\mathrm{O}^0}
   - k_{\mathrm{OH}} N_{\mathrm{O}^+} N_{\mathrm{H}^0} \; ,
\label{eq:dO+dt}
\end{eqnarray}
   where $N_{\mathrm{H}^+}$ is the number density of ionized hydrogen.
The net rate~(\ref{eq:[dz+dt]}) for the ions with a detailed level structure
   (Mg$^{+}$, Fe$^{+}$, and Ba$^{+}$) is calculated including all relevant atomic
   processes.

To close the system of the above rate equations
   (\ref{eq:dN2dt}--\ref{eq:dz+kdt}, \,\ref{eq:dH+dt}--\ref{eq:dH3+dt},
   \,\ref{eq:dz+dt}--\ref{eq:dO+dt}) we use the particle conservation
   for hydrogen
\begin{equation}
   N_{\mathrm{H}^0} + N_{\mathrm{H}^+} + N_{\mathrm{H}^-}
   + 2 N_{\mathrm{H}_2} + 2 N_{\mathrm{H}_2^+}
   + 3 N_{\mathrm{H}_3^+} = \frac{\rho X_{\mathrm{H}}}{m_u A_{\mathrm{H}}}
   \label{eq:numbconsH}
\end{equation}
   and for other elements from He to Ba with three ionization stages
\begin{equation}
   N_{\mathrm{z}^0} + N_{\mathrm{z}^+} + N_{\mathrm{z}^{2+}}
   = \frac{\rho X_{\mathrm{z}}}{m_u A_{\mathrm{z}}} \; ,
   \label{eq:numbconsz}
\end{equation}
   and the charge conservation
\begin{equation}
   N_{\mathrm{H}^+} - N_{\mathrm{H}^-} + N_{\mathrm{H}_2^+} +
   N_{\mathrm{H}_3^+} +
   \sum\limits_{{\mathrm z}={\mathrm{He}}}^{{\mathrm{Ba}}}
   (N_{\mathrm{z}^+} + 2 N_{\mathrm{z}^{2+}}) =
   N_{\mathrm{e}} \; ,
   \label{eq:chargecons}
\end{equation}
   where $X_{\mathrm{z}}$ is the mass fraction of element z, and
   $A_{\mathrm{z}}$ is its atomic weight.
\subsection{Gas energy equation}
\label{sec:model-gseq}
The equation of state of a perfect gas can be written as
\begin{equation}
   P_g = ( N_t + N_{\mathrm{e}} ) k T_{\mathrm{e}} \; ,
\label{eq:pg}
\end{equation}
   where $P_g$ is gas pressure and $N_t$ is the total number density of
   atoms, molecules, and ions of all types.
The specific internal energy of particles (per gram) is the sum of
   the translational, excitation,  and ionization energies:
\begin{equation}
   E_g = \frac 32 \frac{ N_t + N_{\mathrm{e}} }{\rho} kT_{\mathrm{e}} +
   E_{exc} + E_{ion} \; ,
\label{eq:eg}
\end{equation}
   where
\begin{equation}
   E_{exc} = \sum\limits_{{\mathrm z}={\mathrm{H}}}^{{\mathrm{Ba^+}}}
   \sum\limits_{k>1} \epsilon_{\mathrm{z},k} N_{\mathrm{z},k}
\label{eq:eexc}
\end{equation}
   is the excitation energy and
\begin{eqnarray}
   E_{ion} & = &
   I_{\mathrm H}N_{{\mathrm H}^+} - I_{{\mathrm H}^-}N_{{\mathrm H}^-}
   + I_{{\mathrm H}_2}N_{{\mathrm H}_2^+}
   + I_{{\mathrm H}_3}N_{{\mathrm H}_3^+}
   \nonumber \\
   & + &
   \sum\limits_{{\mathrm z}={\mathrm{He}}}^{{\mathrm{Ba}}}
   \left[
   I_{{\mathrm z}^0}N_{{\mathrm z}^+}
   +\left( I_{{\mathrm z}^0}+I_{{\mathrm z}^+}\right)N_{{\mathrm z}^{2+}}
   \right] \; ,
\label{eq:eion}
\end{eqnarray}
   is the ionization energy.
Here $\epsilon_{\mathrm{z},k}$ is the excitation energy of state $k$ for the
   corresponding atoms and ions, and
   $I_{\mathrm H}$, $I_{{\mathrm H}^-}$, $I_{{\mathrm H}_2}$,
   $I_{{\mathrm H}_3}$, $I_{{\mathrm z}^0}$, $I_{{\mathrm z}^+}$ are
   the ionization potentials of the corresponding atoms, molecules, and ions.

The gas energy equation, the first law of thermodynamics for the material,
   including radiative losses in continuum and lines, the Compton cooling,
   and non-thermal heating thus reads
\begin{eqnarray}
   {\partial E_g \over \partial t}
   & + & P_g  {\partial \over \partial t} \left( {1 \over \rho} \right)
   = \frac{4\pi}{\rho} \int_{0}^\infty
   \left( \kappa_\nu J_\nu - \eta_\nu^t - \eta_\nu^{2q} \right) d\nu
   \nonumber \\
   & - &
   \frac{4\pi}{\rho}
   \left(
   \sum\limits_{{\mathrm z}={\mathrm{H}}}^{{\mathrm{Ba^+}}}
   \sum\limits_{l,u} \eta_{ul} + \eta^{\mathrm C} \right)
   + \varepsilon^\gamma \; ,
\label{eq:Egas}
\end{eqnarray}
   where
\begin{equation}
   \eta^{\mathrm C} = \frac {4 k T_{\mathrm e}}{m_{\mathrm e} c^2}
   \sigma_{\mathrm e} N_{\mathrm e} J
   \left( 1 - \frac {T_{\mathrm r}}{T_{\mathrm e}} \right)
\label{eq:etac}
\end{equation}
   is the net emissivity of the Compton cooling (Weymann \cite{weymann66}).
Here $\kappa_\nu$ and $\eta_\nu^t$ are the monochromatic true absorption
   coefficient corrected for stimulated emission and the monochromatic
   thermal emissivity, including bound-free and free-free processes for
   all atoms and ions, respectively (Mihalas \cite{mihalas78});
   $\eta_\nu^{2q}$ is the two-photon emissivity  of hydrogen;
   $\eta_{ul}$ is the net frequency-averaged emissivity of
   line~(\ref{eq:etaul});
   $\varepsilon^\gamma$ is the rate of energy deposition from
   the radioactive decays;
   $\sigma_{\mathrm e}$ is the Thomson scattering cross section;
   $J$ is the integrated mean intensity of continuum;
   $T_{\mathrm r}=(\pi J / \sigma_{\mathrm R})^{1/4}$ is the radiation
   temperature and $\sigma_{\mathrm R}$ is the Stefan-Boltzmann constant.
The energy equation determines the electron temperature in the SN atmosphere
   which is the same for neutrals and ions.

\subsection{Computational method}
\label{sec:model-method}

A hydrodynamic model with 300 mass zones provides us the velocity $v(r,t)$
   and density $\rho(r,t)$ profiles, the photospheric radius $R_\mathrm{ph}(t)$
   and effective temperature $T_\mathrm{eff}(t)$.
The simplified description of the continuum radiation field as diluted
   photospheric radiation permits us to solve for a given mass zone the
   rate equations
   (\ref{eq:dN2dt}--\ref{eq:dz+kdt}, \,\ref{eq:dH+dt}--\ref{eq:dH3+dt},
   \,\ref{eq:dz+dt}--\ref{eq:dO+dt}), the particle conservation equations
   (\ref{eq:numbconsH}, \,\ref{eq:numbconsz}), the charge conservation
   equation (\ref{eq:chargecons}), and the gas energy equation (\ref{eq:Egas})
   independently of other mass zones.

As the supernova envelope expands the photosphere moves inward the envelope.
In a steady state the physical values in the atmosphere at certain moment
   $t$ are determined by the instant density and radiation field.
In a time-dependent approach one has to start from a moment
   $t_\mathrm{ph}$ at which a given mass zone crosses the photosphere, and
   then to follow the time evolution until $t$.
To obtain the distribution of physical values in the atmosphere this
   procedure should be applied to each mass zone.

The simultaneous solution of the above equations
   (\ref{eq:dN2dt}--\ref{eq:dNkdt},
   \,\ref{eq:dH+dt}--\ref{eq:dH3+dt}, \,\ref{eq:dz+dt}--\ref{eq:chargecons},
   \,\ref{eq:Egas}) is achieved by reducing them to a system of ordinary
   differential equations.
In particular, the gas energy equation (\ref{eq:Egas}) is reduced to the
   ordinary differential equation for electron temperature $T_{\mathrm e}$.
To properly formulate the problem of time evolution of physical values for
   a given mass zone, initial data must be specified at the photospheric level.
This is done by solving the equations
   (\ref{eq:dN2dt}--\ref{eq:dNkdt}, \,\ref{eq:dH+dt}--\ref{eq:dH3+dt},
   \,\ref{eq:dz+dt}--\ref{eq:chargecons}) at the photospheric level
   assuming the steady state for a given electron temperature $T_{\mathrm e}$.
This is the simplest way to model the non-equilibrium conditions at the SN
   photosphere.
The derived system of ordinary differential equations is stiff and
   is integrated by the implicit method of Gear (\cite{gear71})
   with an automatic choice of both the time integration step and
   the order of accuracy of the method.

The recovered time-dependent structure of the atmosphere is used then
   to calculate spectra at selected epochs.
The spectra are modelled by Monte Carlo technique with relativistic effects
   included (Mihalas \cite{mihalas78}; Jeffery \cite{jeffery93},
   \cite{jeffery95}).
We suggest that the photosphere diffusively reflects the incident photons
   (Chugai \& Utrobin \cite{cu00}).
The line scattering is generally non-conservative and is described in terms
   of the line scattering albedo.
The Thomson scattering on free electrons and Rayleigh scattering on neutral
   hydrogen are taken into account.

\section{Results}
\label{sec:result}

\subsection{Early H$\alpha$ and the hydrodynamic model}
\label{sec:res-hahyd}

Preliminary modelling of H$\alpha$ has revealed that at the early epoch
   the H$\alpha$ line strength was sensitive to the density of the outer
   layers.
A similar sensitivity of the H$\alpha$ strength to the density was found
   in the steady-state model (Eastman \& Kirshner \cite{ek89}).
We use therefore the H$\alpha$ on day 4.64 to constrain the kinetic energy
   of the hydrodynamic model.

Assuming the initial electron temperature equal to the effective temperature
   $T_\mathrm{e}=T_\mathrm{eff}$ we solved the system of rate and energy
   equations and then computed H$\alpha$ profile for the low-energy model,
   i.e., for the model N with the kinetic energy $E=10^{51}$ erg
   (Utrobin \cite{utrobin04}).
The corresponding bolometric light curve and the calculated H$\alpha$ profile
   on day 4.64 are shown in Figs.~\ref{fig:hydmods}a and \ref{fig:hydmods}b.
The computed H$\alpha$ line has too weak absorption at high velocities
   compared to the observations (Phillips et al. \cite{phhn88}).
The situation, however, is markedly improved for the high-energy model
   with $E=1.5\times10^{51}$ erg (Figs.~\ref{fig:hydmods}c and
   \ref{fig:hydmods}d).
The larger strength of the H$\alpha$ absorption in the high-energy model
   is the direct outcome of the higher density in the outer layers.
Thus, the strength of the H$\alpha$ absorption at high velocities indicates 
   the favoring energy-to-mass ratio 
   $E/M\approx 0.83\times 10^{50} \mathrm{erg}\,M_{\sun}^{-1}$.
We adopt the model with the high energy as the standard model for our
   computations of the SN~1987A atmosphere.

\subsection{Evolution of hydrogen ionization and H$\alpha$}
\label{sec:res-h}

With the adopted photospheric continuum and the initial electron temperature
   we are able to compute for given moments the distribution of the essential
   physical values in the atmosphere, including the hydrogen level populations.
The initial electron temperature at the photosphere is adopted equal to
   the effective temperature $T_{\mathrm{eff}}$ which is taken from
   the hydrodynamical calculations (Utrobin \cite{utrobin04}).

\subsubsection{Day 4.64}
\label{sec:res-hday4}
\begin{figure}[t]
   \resizebox{\hsize}{!}{\includegraphics{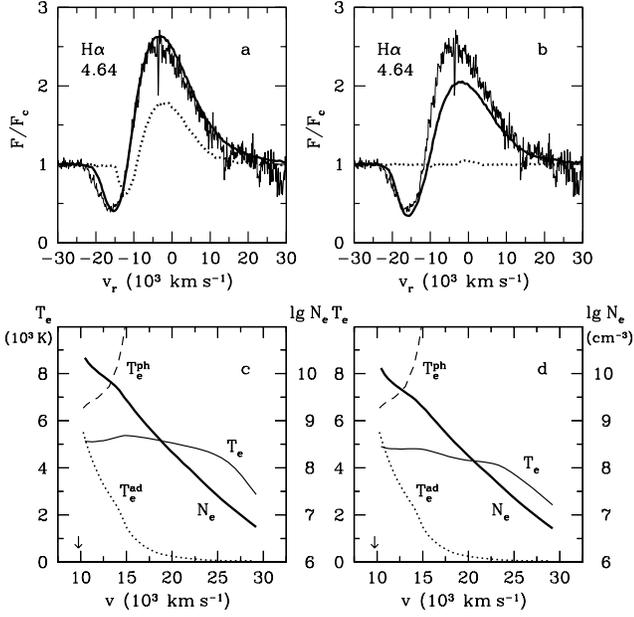}}
   \caption{%
   H$\alpha$ line profile and the structure of the atmosphere on day 4.64.
   Panel (\textbf{a}): H$\alpha$ profile computed with the time-dependent
      model A (\emph{thick solid line\/}) and with the steady-state model A
      (\emph{dotted line\/}), and overplotted on the observed profile
      obtained by Phillips et al. (\cite{phhn88}) (\emph{thin solid line\/}).
   Panel (\textbf{b}): the same as panel (a) but for the model B.
   Panel (\textbf{c}): the calculated electron number density (\emph{thick
      solid line\/}) and electron temperature (\emph{thin solid line\/})
      in the time dependent model A.
      The \emph{dashed line\/} shows the initial electron temperature
      $T_{\mathrm{e}}^{\mathrm{ph}}$ at the photosphere as a function of
      velocity, while the \emph{dotted line\/} is the distribution of
      the electron temperature $T_{\mathrm{e}}^{\mathrm{ad}}$ evaluated
      with the purely adiabatic cooling.
      Arrow indicates the photospheric level.
   Panel (\textbf{d}): the same as panel (c) but for the model B.
   }
   \label{fig:haday4}
\end{figure}
On day 4.64 the H$\alpha$ profile computed in the time-dependent model A
   fits the observed one fairly well (Fig.~\ref{fig:haday4}a), while
   the model B underproduces the emission component (Fig.~\ref{fig:haday4}b).
This suggests that the average intensity of the UV radiation in the atmosphere
   at this and earlier epochs is closer to that of the model A.
The steady-state model A produces rather strong H$\alpha$ line although both
   the absorption and emission components are markedly weaker than in the
   observed line, while in the steady-state model B the H$\alpha$ line is
   extremely weak.
The weakness of the H$\alpha$ line in the steady-state model reflects the
   simple fact that the steady-state ionization is lower than that in the
   time-dependent model.

As expected, the electron temperature in the model A is larger than in the
   model B since the radiative heating rate in the model A with the black-body
   UV continuum is higher than in the model B (Figs.~\ref{fig:haday4}c and
   \ref{fig:haday4}d).
The remarkable property of the distribution of electron temperature in the
   atmosphere for both models is a fast drop from the initial value at the
   photosphere of $T_{\mathrm{eff}}$ to $\sim$ (0.75--0.80) $T_{\mathrm{eff}}$.
This drop results from a simple fact that the electron temperature calculated
   in the diluted radiation field is lower than the effective temperature.
The same physics leads to a well known relation for the boundary temperature
   of gray atmosphere $T_0 \approx 0.81 T_{\mathrm{eff}}$ (see, e.g., Mihalas
   \cite{mihalas78}).

Although the electron temperature is relatively high in the most of the
   atmosphere, $T_{\mathrm{e}}>4000$ K, (Figs.~\ref{fig:haday4}c and
   \ref{fig:haday4}d) it is insufficient to maintain the large optical depth
   of H$\alpha$ demonstrated by the time-dependent models.
The primary mechanism of the population of the second level is
   the recombination to excited levels.

Noteworthy, the populations of excited levels are small numbers that are
   controlled by the relatively fast transition rates.
As a result the relaxation times for the level populations of hydrogen are
   small: $N_{\mathrm{H}^0,k}/(dN_{\mathrm{H}^0,k}/dt) \ll t$.
So the populations of excited levels may be considered in the steady-state
   approximation.
We conclude that the population of the second hydrogen level, and the H$\alpha$
   optical depth as well, is determined mainly by the current values of the
   electron concentration and the photoionization rates, which are essentially
   time-dependent.

\subsubsection{Day 8.69}
\label{sec:res-hday8}
\begin{figure}[t]
   \resizebox{\hsize}{!}{\includegraphics{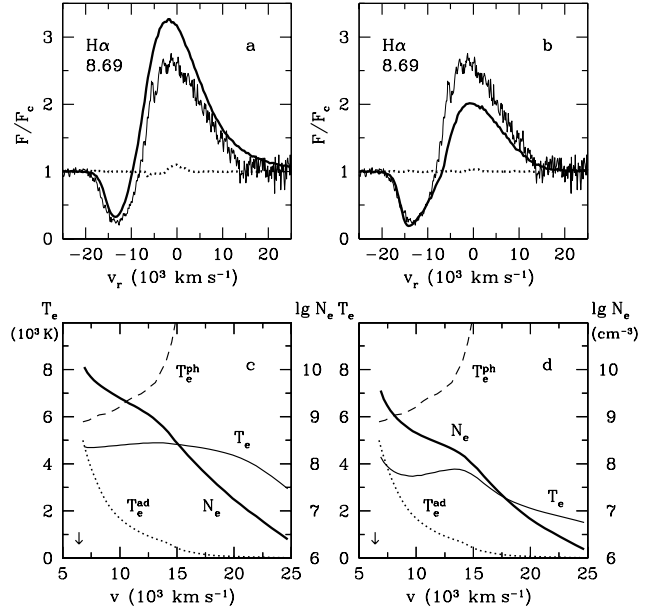}}
   \caption{%
   H$\alpha$ line profile and the structure of the atmosphere on day 8.69.
   See Fig.~\ref{fig:haday4} legend for details.
   }
   \label{fig:haday8}
\end{figure}
On day 8.69 the both A and B models reproduce satisfactorily the absorption
   component of H$\alpha$ (Figs.~\ref{fig:haday8}a and \ref{fig:haday8}b).
The strength of observed emission lies between two models which implies that
   the actual radiation field in the UV band is intermediate between A and B
   cases.
The steady-state approach in both models predicts very weak H$\alpha$ line at
   this epoch.
This is qualitatively consistent with the steady-state modelling of
   Mitchell et al. (\cite{mbb02}).

The electron concentration shows a kink at about 14\,000 km\,s$^{-1}$
   (Figs.~\ref{fig:haday8}c and \ref{fig:haday8}d) that was only barely
   seen at the previous date (Figs.~\ref{fig:haday4}c and \ref{fig:haday4}d).
The position of this kink correlates with the transition from the full
   ionization to partial ionization at the photosphere that occurred
   after $t \approx 2$ days.

\subsubsection{Day 14.68 and 19.68}
\label{sec:res-hday1419}
\begin{figure}[t]
   \resizebox{\hsize}{!}{\includegraphics{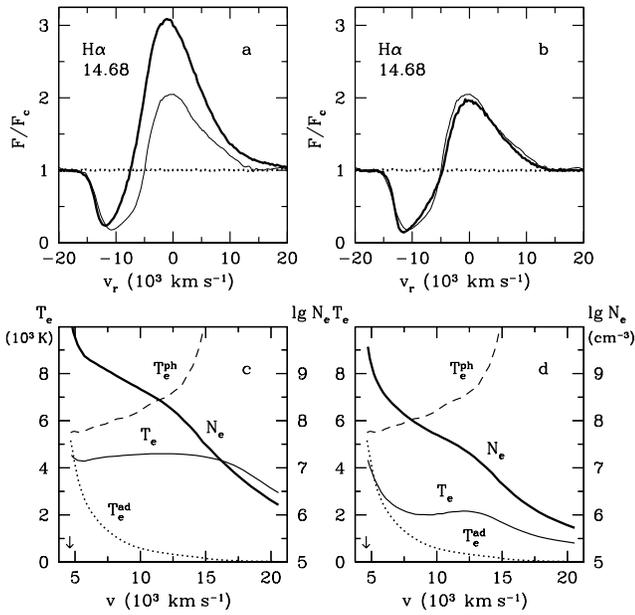}}
   \caption{%
   H$\alpha$ line profile and the structure of the atmosphere on day 14.68.
   Observed profile is given by Hanuschik \& Dachs (\cite{hd88}).
   See Fig.~\ref{fig:haday4} legend for details.
   }
   \label{fig:haday14}
\end{figure}
\begin{figure}[t]
   \resizebox{\hsize}{!}{\includegraphics{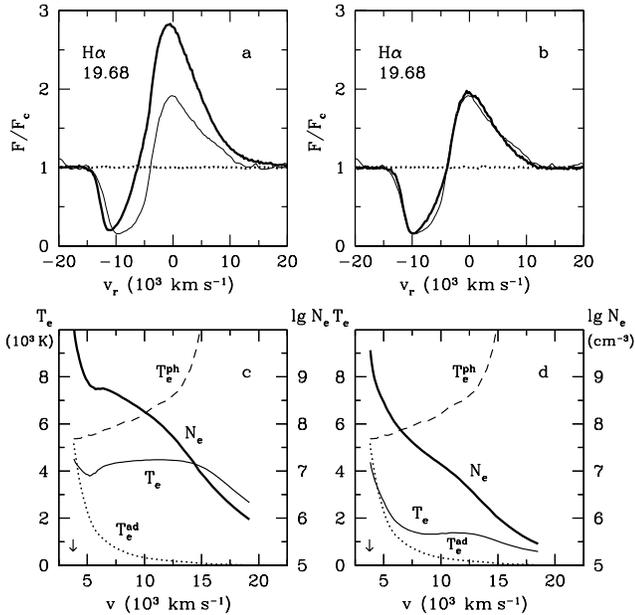}}
   \caption{%
   H$\alpha$ line profile and the structure of the atmosphere on day 19.68.
   Observed profile is given by Hanuschik \& Dachs (\cite{hd88}).
   See Fig.~\ref{fig:haday4} legend for details.
   }
   \label{fig:haday19}
\end{figure}
The calculated H$\alpha$ profiles on day 14.68 (Figs.~\ref{fig:haday14}a and
   \ref{fig:haday14}b) and 19.68 (Figs.~\ref{fig:haday19}a and
   \ref{fig:haday19}b) show similar behavior.
The model B fits to the observed profile fairly well, whereas the emission
   component in the model A is unacceptably strong.
This indicates that around these epochs the UV continuum radiation is closer
   to that of the model B.
The steady-state model at this phase is very weak with practically zero
   absorption and emission components.

At these epochs the adiabatic cooling in the model B dominates at the low
   velocities as indicated by the close run of temperatures $T_{\mathrm{e}}$
   and $T_{\mathrm{e}}^{\mathrm{ad}}$ (Figs.~\ref{fig:haday14}d and
   \ref{fig:haday19}d), and the most of the atmosphere is cooled below
   $\sim 2000$ K.
In the model A the heating due to the photoionization is strong enough to
   maintain rather high temperature $\sim 4000$ K throughout the atmosphere.

\subsubsection{Day 29.68}
\label{sec:res-hday29}
\begin{figure}[t]
   \resizebox{\hsize}{!}{\includegraphics{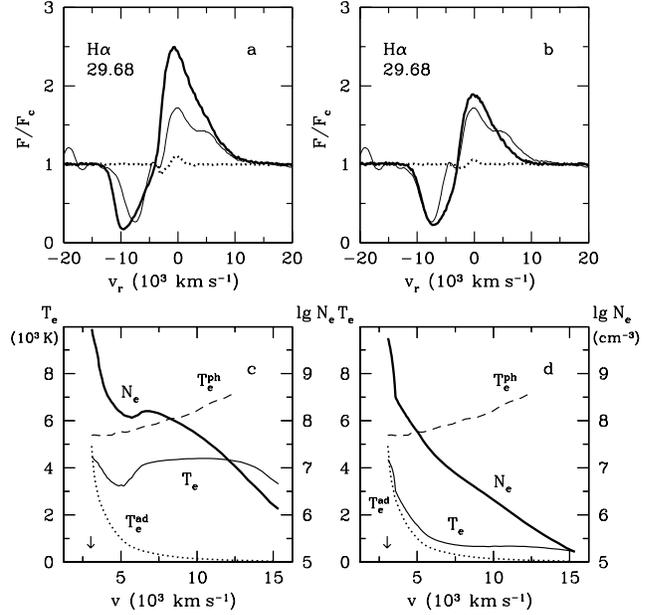}}
   \caption{%
   H$\alpha$ line profile and the structure of the atmosphere on day 29.68.
   Observed profile is given by Hanuschik \& Dachs (\cite{hd88}).
   See Fig.~\ref{fig:haday4} legend for details.
   }
   \label{fig:haday29}
\end{figure}
On day 29.68 the model A leads to the unacceptably strong H$\alpha$ emission
   (Fig.~\ref{fig:haday29}a), while the model B reproduces the overall strength
   of the absorption and emission components of H$\alpha$
   (Fig.~\ref{fig:haday29}b).
The H$\alpha$ line in the steady-state models is extremely weak, although slightly
   stronger than on day 14.68 and 19.68.

The behavior of electron temperature $T_{\mathrm{e}}$ and electron
   concentration $N_{\mathrm{e}}$ is very different for both models
   (Figs.~\ref{fig:haday29}c and \ref{fig:haday29}d) that emphasizes
   the sensitivity of the result to the behavior of the UV radiation
   field in the atmosphere.
Yet, it is highly remarkable that even the time-dependent model B, in which
   the UV continuum intensity is possibly underestimated, reproduces the
   strength of the H$\alpha$ line in the whole range of the radial velocities.
This means that the admixed $^{56}$Ni in the outer layers is not needed
   to account for the H$\alpha$ line in SN~1987A at the age of $t \leq 30$ days.

\begin{figure}[t]
   \resizebox{\hsize}{!}{\includegraphics[clip, trim=0 0 0 278]{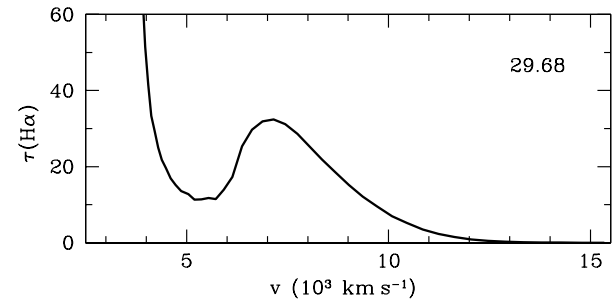}}
   \caption{%
   The H$\alpha$ optical depth on day 29.68 in the model A.
   The local minimum at 5000 km\,s$^{-1}$ is a characteristic
      feature required for the explanation of the blue peak.
   }
   \label{fig:tau29}
\end{figure}
Between day 20 and 29 the additional blue and red peaks emerge in H$\alpha$
   profile, a phenomenon known as "Bochum event" (Hanuschik \& Dachs
   \cite{hd88}).
We believe that the red peak is related with the high velocity $^{56}$Ni
   clump in the far hemisphere, whereas the blue peak reflects a minimum
   in the spherically-symmetric radial distribution of the H$\alpha$ optical
   depth at $\approx 5000$ km\,s$^{-1}$ (Chugai \cite{chugai91b};
   Utrobin et al. \cite{uca95}).
The mechanism responsible for such a minimum is still not identified
   confidently.
Both A and B models do not show any signature of the blue peak in H$\alpha$
   profile.
Remarkably, however, the model A reveals an essentially \emph{non-monotonic}
   behavior of the H$\alpha$ optical depth $\tau(\mbox{H}\alpha)$
   (Fig.~\ref{fig:tau29}) with the position of the minimum at 5000
   km\,s$^{-1}$, just at the place that corresponds to the blue peak.
Unfortunately, the optical depth at this minimum is too high to produce
   the blue peak in the H$\alpha$ profile.
The required value should be 30 times lower.
However the very possibility that the non-monotonic behavior of $\tau(v)$
   may arise in a model with monotonic evolution of the photospheric
   parameters and monotonic density distribution is very promising for
   the explanation of the blue peak in the H$\alpha$ profile at
   the Bochum event phase.

Previously, using the time-dependent model we found that the enhanced hydrogen
   neutralization at about 4000 km\,s$^{-1}$ governed by H$^{-}$ and H$_2^+$
   might be responsible for the local minimum of $N_{\mathrm{e}}$ and thus,
   for the minimum of the H$\alpha$ optical depth $\tau(\mbox{H}\alpha)$
   (Utrobin \& Chugai \cite{uc02}).
However, in that model we adopted rather simple behavior of electron
   temperature $T_{\mathrm{e}}$ in the atmosphere and at the photosphere.
As a result the role of the molecular processes has been significantly 
   exaggerated.

\subsection{Barium problem}
\label{sec:res-ba2}
%
\begin{figure}[t]
   \resizebox{\hsize}{!}{\includegraphics{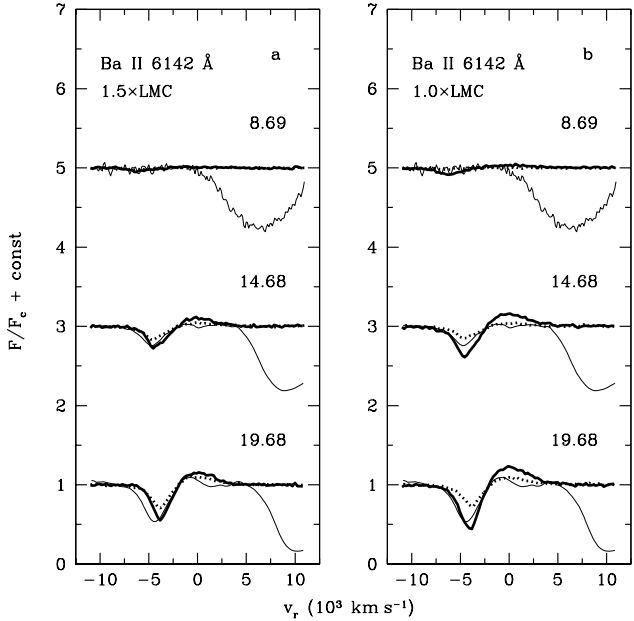}}
   \caption{%
   Evolution of Ba II 6142 \AA\ line from day 8.69 till 19.68 in SN~1987A.
   The Ba II 6142 \AA\ profiles, computed with the time-dependent models
      A (\emph{dotted line\/}) and B (\emph{thick solid line\/}),
      are overplotted on those observed by Phillips et al. (\cite{phhn88})
      and Hanuschik \& Dachs (\cite{hd88}) (\emph{thin solid line\/}).
   Left panel (\textbf{a}): models for the relative Ba overabundance of 1.5.
   Right panel (\textbf{b}): models with the suppressed far UV continuum.
   }
   \label{fig:ba6142}
\end{figure}
The "Ba problem" is the disparity between the high Ba overabundance factor
   $\zeta=$(Ba/Fe)$_{\mathrm{87A}}$/(Ba/Fe)$_{\mathrm{LMC}} \sim 5$
   derived from Ba II 6142 \AA\ line (Mazzali et al. \cite{mlb92})
   and the theoretical predictions of $s$-process calculations.
The $s$-process nucleosynthesis in the massive stars evolving to
   the presupernova star of SN~1987A (Prantzos et al. \cite{phn90}) 
   yields the Ba overabundance of $\zeta \sim$ 1.3--1.4
   in the hydrogen-rich envelope, assuming that a total ejecta mass is 15--18
   $M_{\sun}$, a mass-cut is at 2~$M_{\sun}$ (Woosley \& Weaver \cite{ww95}),
   and the synthesized Ba is completely mixed over the ejecta.
In fact, it is highly difficult to imagine that Ba produced at
   the He-burning stage can be significantly mixed with the hydrogen
   envelope.
More realistic would be a situation with the LMC abundance of Ba in the
   hydrogen envelope.

In our time-dependent approach we recover the distribution of the electron
   concentration that fits to the H$\alpha$ profile on day 14.68 and 19.68
   in the model B.
Therefore we hope that our model can recover a more realistic abundance of
   Ba from Ba II 6142 \AA\ line.
We compute the Ba ionization and excitation, using the multilevel Ba II model
   described in Appendix \ref{apx:atmdt-barium}, for the two choices of the far
   UV ($10.0<h\nu<13.6$ eV) radiation in the SN envelope.
In the first version the continuum in this energy range is the diluted
   photospheric radiation --- our standard assumption for models A and B
   (Fig.~\ref{fig:flux}).
In the second version we follow the suggestion of Mazzali et al. (\cite{mlb92}) 
   that the far UV continuum should be essentially suppressed
   by the line-blocking (in fact Rayleigh scattering also contributed to
   the blocking effect).
This assumption essentially increases the ionization fraction of Ba II and
   thus decreases the overabundance factor.
We, however, modify this assumption making it more realistic: we adopt
   that the far UV ($10.0<h\nu<13.6$ eV) radiation is in equilibrium with matter
   and is characterized by the local electron temperature of our time-dependent
   model.
This is the same approximation as accepted above for $h\nu>13.6$ eV.
Unfortunately, the IUE observations cannot provide a confident verification 
   of our assumption about the far UV flux in SN~1987A because of the 
   dominant contribution of the light from the stars 2 and 3 to this 
   band (Pun et al. \cite{pks95}).
We find that the accepted suppression of the far UV continuum does not affect
   the computed H$\alpha$ line for A and B cases at all.

The Ba II 6142 \AA\ line profile calculated for three phases preceding the
   Bochum event are shown in Fig.~\ref{fig:ba6142}.
In the first version of the continuum the Ba relative abundance of $\zeta=1.5$
   is required to reproduce the Ba II 6142 \AA\ line (Fig.~\ref{fig:ba6142}a),
   while in the second case the LMC abundance of Ba ($\zeta=1$) is quite enough
   (Fig.~\ref{fig:ba6142}b).
We find that for $\zeta=5$ the blue wing of the line can be reproduced in the
   model B on day 14.68 and 19.68 with the photospheric far UV radiation.
Note that Mazzali et al. (\cite{mlb92}) obtained a similar overabundance
   in the model with the essentially suppressed far UV radiation.
In our model with the suppressed far UV radiation the LMC abundance of Ba suffices
   to account for the Ba II 6142 \AA\ line on day 14.68 and 19.68
   (Fig.~\ref{fig:ba6142}b).
Note that in this case the model B produces too strong Ba II 6142 \AA\ line on
   day 8.69, while the model A produces too weak line.
This is consistent with our previous conclusion that at this epoch the H$\alpha$
   line requires the intermediate continuum between model A and B.

We come to the remarkable result: the model B with the suppressed far UV radiation
   that was also assumed in the model of Mazzali et al. (\cite{mlb92})
   recovers the LMC abundance of Ba from the Ba II 6142 \AA\ line.
Therefore, we conclude that at present there is no serious reason anymore to
   suggest the overabundance of Ba in the hydrogen-rich envelope of SN~1987A.
The "Ba problem" in SN~1987A was mainly a by-product of the steady-state
   approximation used for the determination of hydrogen ionization.

\section{Discussion and conclusions}
\label{sec:disc+conc}

Our goal has been to explore the role of time-dependent effects in the
   ionization and excitation of hydrogen in the atmosphere of SN~1987A.
We have developed the model that calculated the time-dependent kinetics of relevant
   processes together with the time-dependent energy balance.
The computations have confirmed our previous conclusion (Utrobin \& Chugai \cite{uc02})
   that the effect of the time-dependent ionization in SN~1987A at the
   photospheric epoch is strong: the hydrogen ionization in the atmosphere is
   significantly larger and, as a result, the hydrogen excitation is higher
   as well.
The H$\alpha$ formation at the photospheric epoch during the first month is
   essentially controlled by the time-dependent ionization effect and the
   additional excitation by the external $^{56}$Ni is not needed to account
   for the strength of H$\alpha$ line.

Although we have failed to reproduce the additional blue peak in the H$\alpha$
   line (Bochum event) that emerged between day 20 and 29 (Hanuschik \& Dachs
   \cite{hd88}), we have found that on day 29.68 one of the models revealed
   a non-monotonic behavior of the H$\alpha$ optical depth with the
   positions of maximum and minimum that were suggested by the H$\alpha$
   profile at this epoch.
This result raises a hope that the blue peak will be explained in a more
   advanced time-dependent model with a correct treatment of the UV radiation
   transfer.

At the epoch preceding the Bochum event the time-dependent model produces
   the realistic distribution of the hydrogen fractional ionization in
   the atmosphere.
With this structure we are able to reproduce the Ba II 6142 \AA\ line
   in SN~1987A on day 14.68 and 19.68 with the LMC abundance of Ba.
We believe, therefore, that the Ba problem, i.e., unacceptably high Ba
   overabundance in SN~1987A, is resolved.
The missing physics of the former models is identified: the time-dependent
   ionization that provides a higher electron concentration in the atmosphere
   and thus more efficient recombination of Ba III into Ba II compared to
   the steady-state models.

\begin{acknowledgements}
V.P.U. is grateful to Wolfgang Hillebrandt and Ewald M\"{u}ller for hospitality
   during staying at the MPA, and also would like to thank Manuel Bautista and
   Keith Butler for kindly providing him with the photoionization cross
   sections of singly ionized iron and barium, respectively.
N.N.C. thanks Roger Chevalier for hospitality in Department of Astronomy of
   the University of Virginia, where a part of this work has been done.
This work has been supported in part by the Russian Foundation for
   Fundamental Research (01-02-16295).
\end{acknowledgements}

\appendix

\section{Atomic data}
\label{apx:atmdt}
\subsection{Hydrogen}
\label{apx:atmdt-hydr}
For hydrogen we use a model atom consisting of 15 levels all of them being
   treated as single levels.
The energy levels and oscillator strengths for the all transitions are taken
   from Wiese et al. (\cite{wsg66}).
The radiative probability of the hydrogen 2s--1s two-photon transition
   and the distribution of the hydrogen two-photon continuum are given by
   Nussbaumer \& Schmutz (\cite{ns84}).
The photoionization cross sections of hydrogen are computed with
   the formulae and tables of Karzas \& Latter (\cite{kl61}).
The temperature-averaged free-free Gaunt factor is interpolated from
   the table and accurate extrapolations given by Sutherland
   (\cite{sutherland98}).
The electron collisional excitation and deexcitation rates for atomic hydrogen
   are calculated with the effective collision strengths taken from
   Scholz et al. (\cite{swbs90}), Callaway (\cite{callaway94}),
   Aggarwal et al. (\cite{abbkp91}), and Giovanardi et al. (\cite{gnp87}).
The electron collisional ionization rates for atomic hydrogen are
   calculated using the approximate formulae of Johnson (\cite{johnson72}).
The rate coefficients for charge transfer in collisions of O$^{+}$ with H and
   H$^{+}$ with O are given by Stancil et al. (\cite{ssk99}).
\subsection{Singly ionized magnesium}
\label{apx:atmdt-magn}
For singly ionized magnesium we use a model ion consisting of 9 single levels.
The energy levels and oscillator strengths for the all transitions are compiled
   from the Atomic Spectroscopic Database (ASD) at the National Institute
   of Standards and Technology (NIST) and from Kurucz \& Bell (\cite{kb95}).
The photoionization cross sections of ionized magnesium are taken from the
   Opacity Project atomic database (Cunto et al. \cite{cmoz93}).
The effective collision strengths for the electron excitation of ionized
   magnesium were computed by Sigut \& Pradhan (\cite{sp95}).
The electron collisional ionization rates for singly ionized magnesium are
   evaluated by fits of Voronov (\cite{voronov97}), the approximate formula
   of Milkey \& Mihalas (\cite{mm74}), and the semi-empirical formula of
   Lotz (\cite{lotz69}).
\subsection{Singly ionized iron}
\label{apx:atmdt-iron}
For the specific case of the Fe II ion a 30-term, multilevel atom model
   is employed.
A total number of 212 transitions between these terms from 2150 \AA\
   to 55 $\mu$m are calculated.
The energy levels and oscillator strengths for the all transitions in
   singly ionized iron are compiled from ASD at NIST and from Kurucz \& Bell
   (\cite{kb95}).
The photoionization cross sections of ionized iron are taken from Nahar \&
   Pradhan (\cite{np94}) and Bautista \& Pradhan (\cite{bp98}), and averaged
   over resonance structures according to Bautista et al. (\cite{brp98}).
The Maxwellian averaged rate coefficients for the infrared, optical and
   ultraviolet transitions in ionized iron are taken from Zhang \& Pradhan
   (\cite{zp95}).
The collision strengths for some transitions are calculated by using an
   empirical formula derived from the available experimental and theoretical
   data for Fe II (Li et al. \cite{lms93}).
The rate of direct collisional ionization for singly ionized iron from the ground
   state is computed using the fits from Voronov (\cite{voronov97}) and
   the electron collisional ionization rates from excited levels are
   evaluated by the semi-empirical formula of Lotz (\cite{lotz69}).
\subsection{Singly ionized barium}
\label{apx:atmdt-barium}
For the case of the Ba II ion a 17-term, multilevel atom model is used with
   a total number of 56 transitions between these terms from 1500 \AA\ to
   58 $\mu$m.
The energy levels and oscillator strengths for the all transitions in
   singly ionized barium are taken from Kurucz \& Bell (\cite{kb95}).
The photoionization cross sections of ionized barium are given by Butler
   (\cite{butler00}) and averaged over resonance structures according to
   Bautista et al. (\cite{brp98}).
The electron collisional excitation rates for ionized barium are taken from
   Sch\"{o}ning \& Butler (\cite{sb98}) and compiled from Sobelman et al.
   (\cite{svy81}) and Van Regemorter (\cite{regemorter62}).
The electron collisional ionization rates are evaluated by the semi-empirical
   formula of Lotz (\cite{lotz69}).
\subsection{Other atoms and ions}
\label{apx:atmdt-other}
The following elements are included in the non-steady study: He, C, N, O, Ne,
   Na, Mg, Si, S, Ar, Ca, Fe, and Ba.
All elements are treated with the three ionization stages.
The atoms and ions are assumed to consist of the ground state and continuum
   except the singly ionized elements with a detailed energy level structure.
Atomic weights and ionization potentials are from Allen (\cite{allen73}).
The partition functions are calculated with the polynomial approximation fit
   obtained by Irwin (\cite{irwin81}).
The photoionization cross sections of atoms and ions are evaluated with data
   of Verner \& Yakovlev (\cite{vy95}) and Verner et al. (\cite{vfky96}).
As for H the temperature-averaged free-free Gaunt factor is taken from
   Sutherland (\cite{sutherland98}).
The free-free absorption coefficient is calculated with the effective nuclear
   charge including screening effects (Sutherland \& Dopita \cite{sd93}).
The electron collisional ionization rates for atoms and ions are computed
   using the approximate formulae of Voronov (\cite{voronov97}).
\section{Molecular data}
\label{apx:moldt}
\subsection{Negative hydrogen ion}
\label{apx:moldt-negat}
The photoionization cross section data for negative hydrogen ion are taken
   from Wishart (\cite{wishart79}).
The free-free absorption coefficient of negative hydrogen ion was calculated
   by Bell \& Berrington (\cite{bb87}).
\subsection{Other hydrogen molecules}
\label{apx:moldt-other}
The neutral-neutral reactions for hydrogen molecules are listed in
   Table~\ref{ListNNReac} with their names and references from which
   the corresponding rate coefficients are taken.
\begin{table}
   \caption[]{List of neutral-neutral reactions.}
   \label{ListNNReac}
   \centering
   \begin{tabular}{p{0.03\linewidth}p{0.43\linewidth}p{0.42\linewidth}}
   \hline\hline
   \noalign{\smallskip}
   No. & Reaction & Source \\
   \noalign{\smallskip}
   \hline
   \noalign{\smallskip}
   1 & H + H $\to$ H$^{+}$ + H + e$^{-}$ & Hollenbach \& McKee \cite{hm89} \\
   2 & H + H $\to$ H$^{+}$ + H$^{-}$ & Hollenbach \& McKee \cite{hm89} \\
   3 & H + H + H $\to$ H$_{2}$ + H & Palla et al. \cite{pss83} \\
   4 & H + H + H$_{2}$ $\to$ 2H$_{2}$ & Palla et al. \cite{pss83} \\
   5 & H(2s) + H $\to$ H$^{+}_{2}$ + e$^{-}$ & Rawlings et al. \cite{rdb93} \\
   6 & H(2p) + H $\to$ H$^{+}_{2}$ + e$^{-}$ & Rawlings et al. \cite{rdb93} \\
   7 & H(3) + H $\to$ H$^{+}_{2}$ + e$^{-}$ & Rawlings et al. \cite{rdb93} \\
   8 & H(2s) + H(1s) $\to$ H$_{2}$ + $\gamma$ & Latter \& Black \cite{lb91} \\
   9 & H(2p) + H(1s) $\to$ H$_{2}$ + $\gamma$ & Latter \& Black \cite{lb91} \\
  10 & H + H$_{2}$ $\to$ 3H & Roberge \& Dalgarno \cite{rd82} \\
  11 & H + H$_{2}$ $\to$ H$^{-}$ + H$^{+}_{2}$ & Hollenbach \& McKee \cite{hm89} \\
  12 & H + H$_{2}$ $\to$ H$^{+}_{2}$ + H + e$^{-}$ & Hollenbach \& McKee \cite{hm89} \\
  13 & H + H$_{2}$ $\to$ H$^{+}$ + H$_{2}$ + e$^{-}$ & Hollenbach \& McKee \cite{hm89} \\
  14 & H$^{*}$ + H$_{2}$ $\to$ H$^{+}_{3}$ + e$^{-}$ & Culhane \& McCray \cite{cm95} \\
  15 & H$_{2}$ + H$_{2}$ $\to$ H$^{+}_{2}$ + H$_{2}$ + e$^{-}$ & Hollenbach \& McKee \cite{hm89} \\
  16 & H$_{2}$ + H$_{2}$ $\to$ H$_{2}$ + 2H & Rawlings \cite{rawlings88} \\
   \noalign{\smallskip}
   \hline
   \end{tabular}
\end{table}
Tables~\ref{ListINReac} and \ref{ListIIReac} list the ion-neutral and ion-ion
   reactions for hydrogen molecules with the relevant references, respectively.
\begin{table}
   \caption[]{List of ion-neutral reactions.}
   \label{ListINReac}
   \centering
   \begin{tabular}{p{0.03\linewidth}p{0.43\linewidth}p{0.42\linewidth}}
   \hline\hline
   \noalign{\smallskip}
   No. & Reaction & Source \\
   \noalign{\smallskip}
   \hline
   \noalign{\smallskip}
   17 & H$^{-}$ + H $\to$ 2H + e$^{-}$ & Abel et al. \cite{aazn97} \\
   18 & H$^{-}$ + H $\to$ H$_{2}$ + e$^{-}$ & Galli \& Palla \cite{gp98} \\
   19 & H$^{+}$ + H $\to$ H$^{+}_{2}$ + $\gamma$ & Galli \& Palla \cite{gp98} \\
   20 & H$^{+}$ + H$_{2}$ $\to$ H$^{+}_{3}$ + $\gamma$ & Galli \& Palla \cite{gp98} \\
   21 & H$^{+}$ + H$_{2}$ $\to$ H$^{+}_{2}$ + H & Galli \& Palla \cite{gp98} \\
   22 & H$^{+}_{2}$ + H $\to$ H$_{2}$ + H$^{+}$ & Galli \& Palla \cite{gp98} \\
   23 & H$^{+}_{2}$ + H$_{2}$ $\to$ H$^{+}_{3}$ + H & Hollenbach \& McKee \cite{hm89} \\
   24 & H$^{+}_{2}$ + H$_{2}$ $\to$ H$_{2}$ + H$^{+}$ + H & Hollenbach \& McKee \cite{hm89} \\
   25 & H$^{+}_{3}$ + H $\to$ H$_{2}$ + H$^{+}_{2}$ & Galli \& Palla \cite{gp98} \\
   26 & H$^{+}_{3}$ + H$_{2}$ $\to$ H$_{2}$ + H$^{+}_{2}$ + H & Hollenbach \& McKee \cite{hm89} \\
   27 & H$^{+}_{3}$ + H$_{2}$ $\to$ 2H$_{2}$ + H$^{+}$ & Hollenbach \& McKee \cite{hm89} \\
   \noalign{\smallskip}
   \hline
   \end{tabular}
\end{table}
\begin{table}
   \caption[]{List of ion-ion reactions.}
   \label{ListIIReac}
   \centering
   \begin{tabular}{p{0.03\linewidth}p{0.43\linewidth}p{0.42\linewidth}}
   \hline\hline
   \noalign{\smallskip}
   No. & Reaction & Source \\
   \noalign{\smallskip}
   \hline
   \noalign{\smallskip}
   28 & H$^{-}$ + H$^{+}$ $\to$ 2H & Galli \& Palla \cite{gp98} \\
   29 & H$^{-}$ + H$^{+}$ $\to$ H$^{+}_{2}$ + e$^{-}$ & Galli \& Palla \cite{gp98} \\
   30 & H$^{-}$ + H$^{+}_{2}$ $\to$ H + H$_{2}$ & Abel et al. \cite{aazn97} \\
   \noalign{\smallskip}
   \hline
   \end{tabular}
\end{table}
The electron reactions for hydrogen molecules are listed in
   Table~\ref{ListELReac} with their names and references from which
   the corresponding rate coefficients are taken.
\begin{table}
   \caption[]{List of electron reactions.}
   \label{ListELReac}
   \centering
   \begin{tabular}{p{0.03\linewidth}p{0.43\linewidth}p{0.42\linewidth}}
   \hline\hline
   \noalign{\smallskip}
   No. & Reaction & Source \\
   \noalign{\smallskip}
   \hline
   \noalign{\smallskip}
   31 & e$^{-}$ + H$^{-}$ $\to$ H + 2e$^{-}$ & Abel et al. \cite{aazn97} \\
   32 & e$^{-}$ + H$_{2}$ $\to$ 2H + e$^{-}$ & Stibbe \& Tennyson \cite{st99} \\
   33 & e$^{-}$ + H$_{2}$ $\to$ H + H$^{-}$ & Fuller \& Couchman \cite{fc00} \\
   34 & e$^{-}$ + H$_{2}$ $\to$ H$^{+}_{2}$ + 2e$^{-}$ & Hollenbach \& McKee \cite{hm89} \\
   35 & e$^{-}$ + H$^{+}_{2}$ $\to$ 2H & Galli \& Palla \cite{gp98} \\
   36 & e$^{-}$ + H$^{+}_{3}$ $\to$ H + H$_{2}$ & Galli \& Palla \cite{gp98} \\
   37 & e$^{-}$ + H$^{+}_{3}$ $\to$ 3H & Hollenbach \& McKee \cite{hm89} \\
   \noalign{\smallskip}
   \hline
   \end{tabular}
\end{table}
Table~\ref{ListPHProc} gives the photoionization and photodissociation
   processes, their names, and references from which the photoionization
   and photodissociation cross sections are taken.
These cross sections are used in calculating the ionization and dissociation
   rates in the radiation field presented in Sect.~\ref{sec:model-cont}.
\begin{table}
   \caption[]{List of photoionization and photodissociation processes.}
   \label{ListPHProc}
   \centering
   \begin{tabular}{p{0.03\linewidth}p{0.43\linewidth}p{0.42\linewidth}}
   \hline\hline
   \noalign{\smallskip}
   No. & Process & Source \\
   \noalign{\smallskip}
   \hline
   \noalign{\smallskip}
   1 & H$_{2}$ + $\gamma$ $\to$ H$^{+}_{2}$ + e$^{-}$ & Yan et al. \cite{ysd98}, \cite{ysd01} \\
   2 & H$_{2}$ + $\gamma$ $\to$ 2H & Abel et al. \cite{aazn97} \\
   3 & H$_{2}$ + $\gamma$ $\to$ H$^{*}_{2}$ $\to$ 2H & Abel et al. \cite{aazn97} \\
   4 & H$^{+}_{2}$ + $\gamma$ $\to$ H + H$^{+}$ & Stancil \cite{stancil94} \\
   5 & H$^{+}_{2}$ + $\gamma$ $\to$ 2H$^{+}$ + e$^{-}$ & Shapiro \& Kang \cite{sk87} \\
   \noalign{\smallskip}
   \hline
   \end{tabular}
\end{table}
%
\section{Rate equations for H$^{+}$, H$^{-}$, H$_{2}$, H$_{2}^{+}$,
            and H$_{3}^{+}$}
\label{apx:rteqmolh}
Using the physical values and quantities from Sect.~\ref{sec:model-rteq}
   the rate equation for ionized hydrogen may be written as
\begin{eqnarray}
   {\partial N_{\mathrm{H}^+} \over \partial t}
   & = &
   - {N_{\mathrm{H}^+} \over t_{exp}}
   + \sum\limits_{k=1}^{15} R_{k,c}^\gamma
   \nonumber \\
   & + &
   \sum\limits_{k=1}^{15} N_{\mathrm{H}^0,k} P_k
   - N_{\mathrm{H}^+} N_\mathrm{e} \sum\limits_{k=2}^{15} \alpha_k
   \nonumber \\
   & + &
   N_\mathrm{e} \sum\limits_{k=1}^{15} N_{\mathrm{H}^0,k} q_{k,c}
   - N_{\mathrm{H}^+} N_\mathrm{e}^2 \sum\limits_{k=1}^{15} q_{c,k}
   \nonumber \\
   & + &
   (k_{1} + k_{2}) N_{\mathrm{H}^0}^2
   + k_{13} N_{\mathrm{H}^0} N_{\mathrm{H}_2}
   \nonumber \\
   & + &
   k_{22} N_{\mathrm{H}_2^+} N_{\mathrm{H}^0}
   + k_{24} N_{\mathrm{H}_2^+} N_{\mathrm{H}_2}
   + k_{27} N_{\mathrm{H}_3^+} N_{\mathrm{H}_2}
   \nonumber \\
   & + &
   k_{\mathrm{OH}} N_{\mathrm{O}^+} N_{\mathrm{H}^0}
   + (R_{4} + 2 R_{5}) N_{\mathrm{H}_2^+}
   \nonumber \\
   & - &
   k_{19} N_{\mathrm{H}^+} N_{\mathrm{H}^0}
   - (k_{20} + k_{21}) N_{\mathrm{H}^+} N_{\mathrm{H}_2}
   \nonumber \\
   & - &
   (k_{28} + k_{29}) N_{\mathrm{H}^-} N_{\mathrm{H}^+}
   - k_{\mathrm{HO}} N_{\mathrm{H}^+} N_{\mathrm{O}^0} \; ,
\label{eq:dH+dt}
\end{eqnarray}
   where $R_{4}$ and $R_{5}$ are the photodissociation and dissociative
   photoionization rates evaluated with the relevant cross sections
   from Table~\ref{ListPHProc}, respectively;
   the rate coefficients for the involved reactions are taken from
   Tables~\ref{ListNNReac}--\ref{ListIIReac}.
Radiative recombinations to the ground state of hydrogen lead to immediate
   reionizations from the ground state, so they are omitted.
The rate equation for negative hydrogen ion is
\begin{eqnarray}
   {\partial N_{\mathrm{H}^-} \over \partial t}
   & = &
   - {N_{\mathrm{H}^-} \over t_{exp}}
   + N_{\mathrm{H}^0} N_\mathrm{e} \alpha_{\mathrm{H}^-}
   - N_{\mathrm{H}^-} P_{\mathrm{H}^-}
   \nonumber \\
   & + &
   k_{2} N_{\mathrm{H}^0}^2
   + k_{11} N_{\mathrm{H}^0} N_{\mathrm{H}_2}
   + k_{33} N_{\mathrm{e}} N_{\mathrm{H}_2}
   \nonumber \\
   & - &
   (k_{17} + k_{18}) N_{\mathrm{H}^-} N_{\mathrm{H}^0}
   - k_{30} N_{\mathrm{H}^-} N_{\mathrm{H}_2^+}
   \nonumber \\
   & - &
   (k_{28} + k_{29}) N_{\mathrm{H}^-} N_{\mathrm{H}^+}
   - k_{31} N_{\mathrm{e}} N_{\mathrm{H}^-} \; ,
\label{eq:dH-dt}
\end{eqnarray}
   where $\alpha_{\mathrm{H}^-}$ and $P_{\mathrm{H}^-}$ are the total
   coefficient of photo-attachment of neutral hydrogen and electron to
   negative hydrogen ion and the photo-detachment rate of negative hydrogen
   ion in the continuum radiation field, respectively;
   the rate coefficients for the corresponding reactions are taken from
   Tables~\ref{ListNNReac}--\ref{ListELReac}.
The rate equation for molecular hydrogen H$_{2}$ is given by
\begin{eqnarray}
   {\partial N_{\mathrm{H}_2} \over \partial t}
   & = &
   - {N_{\mathrm{H}_2} \over t_{exp}}
   - (R_{1} + R_{2} + R_{3}) N_{\mathrm{H}_2}
   \nonumber \\
   & + &
   k_{3} N_{\mathrm{H}^0}^3
   + k_{4} N_{\mathrm{H}^0}^2 N_{\mathrm{H}_2}
   \nonumber \\
   & + &
   k_{8} N_{\mathrm{H}^0,1} N_{\mathrm{H}^0,2s}
   + k_{9} N_{\mathrm{H}^0,1} N_{\mathrm{H}^0,2p}
   \nonumber \\
   & + &
   k_{18} N_{\mathrm{H}^-} N_{\mathrm{H}^0}
   + k_{22} N_{\mathrm{H}_2^+} N_{\mathrm{H}^0}
   + k_{25} N_{\mathrm{H}_3^+} N_{\mathrm{H}^0}
   \nonumber \\
   & + &
   k_{27} N_{\mathrm{H}_3^+} N_{\mathrm{H}_2}
   + k_{30} N_{\mathrm{H}^-} N_{\mathrm{H}_2^+}
   + k_{36} N_{\mathrm{e}} N_{\mathrm{H}_3^+}
   \nonumber \\
   & - &
   (k_{10} + k_{11} + k_{12}) N_{\mathrm{H}^0} N_{\mathrm{H}_2}
   \nonumber \\
   & - &
   k_{14} N_{\mathrm{H}_2} N_{\mathrm{H}^0}
   - (k_{15} + k_{16}) N_{\mathrm{H}_2}^2
   \nonumber \\
   & - &
   (k_{20} + k_{21}) N_{\mathrm{H}^+} N_{\mathrm{H}_2}
   - k_{23} N_{\mathrm{H}_2^+} N_{\mathrm{H}_2}
   \nonumber \\
   & - &
   (k_{32} + k_{33} + k_{34}) N_{\mathrm{e}} N_{\mathrm{H}_2} \; ,
\label{eq:dH2dt}
\end{eqnarray}
   where $R_{1}$, $R_{2}$, and $R_{3}$ are
   the photoionization, direct photodissociation, and two-step
   photodissociation (the Solomon process) rates of molecular hydrogen H$_{2}$
   evaluated with the cross sections from Table~\ref{ListPHProc}, respectively;
   the rate coefficients for the involved reactions are taken from
   Tables~\ref{ListNNReac}--\ref{ListELReac}.
Finally, the rate equations for molecular hydrogen H$_{2}^{+}$ and H$_{3}^{+}$
   are
\begin{eqnarray}
   {\partial N_{\mathrm{H}_2^+} \over \partial t}
   & = &
   - {N_{\mathrm{H}_2^+} \over t_{exp}}
   + R_{1} N_{\mathrm{H}_2} - (R_{4} + R_{5}) N_{\mathrm{H}_2^+}
   \nonumber \\
   & + &
   (k_{5} N_{\mathrm{H}^0,2s} + k_{6} N_{\mathrm{H}^0,2p}
   + k_{7} N_{\mathrm{H}^0,3}) N_{\mathrm{H}^0}
   \nonumber \\
   & + &
   (k_{11} + k_{12}) N_{\mathrm{H}^0} N_{\mathrm{H}_2}
   + k_{15} N_{\mathrm{H}_2}^2
   \nonumber \\
   & + &
   k_{19} N_{\mathrm{H}^+} N_{\mathrm{H}^0}
   + k_{21} N_{\mathrm{H}^+} N_{\mathrm{H}_2}
   + k_{25} N_{\mathrm{H}_3^+} N_{\mathrm{H}^0}
   \nonumber \\
   & + &
   k_{26} N_{\mathrm{H}_3^+} N_{\mathrm{H}_2}
   + k_{29} N_{\mathrm{H}^-} N_{\mathrm{H}^+}
   + k_{34} N_{\mathrm{e}} N_{\mathrm{H}_2}
   \nonumber \\
   & - &
   k_{22} N_{\mathrm{H}_2^+} N_{\mathrm{H}^0}
   - (k_{23} + k_{24}) N_{\mathrm{H}_2^+} N_{\mathrm{H}_2}
   \nonumber \\
   & - &
   k_{30} N_{\mathrm{H}^-} N_{\mathrm{H}_2^+}
   - k_{35} N_{\mathrm{e}} N_{\mathrm{H}_2^+}
\label{eq:dH2+dt}
\end{eqnarray}
   and
\begin{eqnarray}
   {\partial N_{\mathrm{H}_3^+} \over \partial t}
   & = &
   - {N_{\mathrm{H}_3^+} \over t_{exp}}
   + k_{14} N_{\mathrm{H}_2} N_{\mathrm{H}^0}
   \nonumber \\
   & + &
   k_{20} N_{\mathrm{H}^+} N_{\mathrm{H}_2}
   + k_{23} N_{\mathrm{H}_2^+} N_{\mathrm{H}_2}
   \nonumber \\
   & - &
   k_{25} N_{\mathrm{H}_3^+} N_{\mathrm{H}^0}
   - (k_{26} + k_{27}) N_{\mathrm{H}_3^+} N_{\mathrm{H}_2}
   \nonumber \\
   & - &
   (k_{36} + k_{37}) N_{\mathrm{e}} N_{\mathrm{H}_3^+} \; ,
\label{eq:dH3+dt}
\end{eqnarray}
   respectively.

\end{document}